\documentclass[allclo,superscriptaddress,eqsecnum,amsfonts,showpacs]{revtex4}
\usepackage{epsfig}

\newcommand{\be}{\begin{equation}}
\newcommand{\ee}{\end{equation}}
\newcommand{\bea}{\begin{eqnarray}}
\newcommand{\eea}{\end{eqnarray}}

\newcommand{\ep}{i\varepsilon}
\newcommand{\nn}{\nonumber}

\begin{document}

\preprint{ \parbox{1.5in}{\leftline{hep-th/??????}}}

\title{Chiral Symmetry Breaking and confinement in Minkowski space QED2+1}

\author{Vladimir ~\v{S}auli}
\affiliation{CFTP and Dept. of Phys.,
IST, Av. Rovisco Pais, 1049-001 Lisbon,
Portugal }
\affiliation{Dept. of Theor. Phys., INP, \v{R}e\v{z} near Prague, AV\v{C}R}

\author{ Zoltan Batiz}
\affiliation{CFTP and Dept. of Phys.,
IST, Av. Rovisco Pais, 1049-001 Lisbon,
Portugal }

\begin{abstract}
Without any analytical assumption we solve the ladder QED2+1 in Minkowski space.  
Obtained complex fermion propagator exhibits confinement in  the sense that it  
has no  pole. Further, we transform Greens functions  
to the Temporal Euclidean space, wherein we  show that in the special case of ladder 
QED2+1 the solution  is fully equivalent to the  Minkowski one. Obvious invalidity of Wick rotation is briefly discussed. The infrared value of the dynamical mass is compared with 
other known approaches, e.g. with the standard Euclidean calculation.

\end{abstract}

\maketitle
%

\section{Introduction}

 Quark confinement in Quantum Chromodynamics (QCD) is a phenomenon of current
interest. Due to the fact that QCD is not easily tractable, various toy models which exhibit QCD low energy phenomena --the confinement and chiral symmetry breaking-- are often investigated.
 These similarities with QCD in the
usual 3+1d Minkowski space has been appreciated in  2+1d Quantum Electrodynamics (QED2+1) time ago \cite{PIS1984,ABKW1986,APNAWI1988}.  
More specifically, based on the Euclidean space study of QED3, the chiral symmetry breaking for a small number flavors has been proposed  for the first time in \cite{APNAWI1988}. Since the scale of dynamical chiral symmetry breaking, being characterized  by a fermion mass in the infrared -$M(0)$-, is
one order of magnitude smaller then the topological dimensioned coupling $e^2$, the 
Schwinger-Dyson equations (SDEs) provide a unique powerful framework for the nonperturbative study, see e.g. most recent studies \cite{BARA2007,BARACORO2008}. 
The importance of the unquenching effect in QED3 for an increasing $N_f$ has been recognized a long time ago \cite{DAKOKO1989,BUPRPRO1992} and reinvestigated in SDEs framework lately \cite{GUHARE1996,MARIS1995,FIALDAMA2004,BARA2007}. Particularly, confinement in relation with  dynamical  complex pole generation in fermion propagator has been discussed in 
\cite{MARIS1995}.  Further analogy with QCD in finite temperature and chemical potential has been explored   
\cite{HEFESUZO2007}. It is noteworthy that QED3 is of current interest as an effective theory of high-temperature superconductors \cite{FRTEVA2002,HERBUT2002,ASTE2005,THHA2007} , Mott insulator \cite{NOHA2005} and graphene \cite{NOVO2005,GUSHCA2007}.

However, the all aforementioned  studies have been done in the
standard Euclidean
space. That is, after performing the standard Wick rotation \cite{WICK} of the timelike
components of the momentum variables (internal integral momentum as well as external one, explicitly $p_3^{E}=-ip_0^{M}$,  the measure $id^3p^{M}=-d^3p^{E}$). It is assumed and widely believed, that the Green`s functions for timelike arguments can be obtained after analytical continuation of the Schwinger functions calculated in  Euclidean space, whilst it is supposed the Euclidean solution itself  represent the correct Minkowski solution for the spacelike arguments. In fact it is more common to view Euclidean space as the definitive frame, see for instance   Sections 2.3 and 6.3 in  \cite{ROBERTS} and 2.1 in \cite{RS2000}.
Therefore, to shed a new light and for the first time,  we solve fermion SDE directly in 2+1 Minkowski space. For this purpose the  ladder approximation of electron SDE is introduced in the  Section II. 

In the  Section III, by using hyperbolic coordinates, we 
solve the SDE directly in the Minkowski space. While for timelike $p^2$ it provides selfconsistent gap equation, i.e. the integral equation with the mass function of the same arguments in both sides of an appropriate gap equation,   
whilst for spacelike Minkowski subspace the solution is naturally constructed from necessarily  known timelike   solution. First numerical results are presented in the Section IV.
Furthermore, we derive the ladder SDE in the Temporal Euclidean (TE) space in the Section V and prove that this exactly leads to the  original  Minkowski formulation for timelike momenta. Recall here,  TE  space  metric is obtained from Minkowski one by the multidimensional  Wick rotations, but now instead for the time component, it is  made for  all the space coordinates of the Lorentz three vector. Usage of  TE space instead of more standard spacelike Euclidean one was suggested first time in \cite{SABA2009b,SAZO2008} as a powerful method when approximating  above pronounced nonperturbative phenomena in true Minkowski space. 

Further comparison (e.g. with perturbation theory) of obtained numerical results for various ratio of the coupling and the electron mass are presented and discussed in Sections IV and V.   The resulting fermion propagator violates reflection positivity- it does not satisfy Khallen-Lehmann representation-  and it has no pole singularity in the timelike region. The most striking result is that the imaginary part of the infrared mass function is automatically generated for a coupling strong enough. The associated quantum -electron living on the plane- can never be on-shell and thus never observed as a free particle, in other words it is permanently confined \cite{CORN1980,GOMA1989,GRIBOV91,ROWIKR1992,MARIS1995}.
 We further discuss the (in-)validity of standard Wick rotation and  conclude in the Section VI.

\section{Fermion SDE in QED2+1, QED3 in ladder approximation}

In our study we employ Minkowski metric $g_{\mu\nu}=diag(1,-1,-1)$, in order to  properly describe chiral symmetry, we use the standard four dimensional Dirac matrices such that they anticomutation relation is  $\left\{\gamma_{\mu},\gamma_{\nu}\right\}=2g_{\mu\nu}$. With these conventions the inverse of the full fermion propagator reads
\bea \label{gap}
S^{-1}(p)&=&\not p- m-\Sigma(p)\, ,
\nn \\
\Sigma(p)&=&ie^2\int \frac{d^3k}{(2\pi)^3}  G^{\mu\nu}(k-p)\Gamma_{\mu}(k,p)S(k)\gamma_{\nu}\, .
\eea

We consider the explicit chiral symmetry breaking  mass term of the form $m\bar{\psi}\psi$
so  parity is conserved. In this case the dressed fermion propagator can be parametrized
by two scalar function like
\be
S(p)=S_v(p)\not p+S_s(p)=\frac{1}{\not p A(p)-B(p)}\, . 
\ee

The full photon propagator $G$ and the electron-positron-photon vertex $\Gamma$ satisfy their own
SDEs (for their general forms see \cite{ROBERTS}).   

The ladder approximation is the simplest selfconsistent approximation which approximate  the unknown Greens functions by their free counterpartners, i.e. $\Gamma_{\mu}=\gamma_{\mu}$ and the photon propagator in linear covariant gauges is
\be
G_{\mu\nu}=\frac{-g_{\mu\nu}+(1-\xi)\frac{k_{\mu}k_{\nu}}{k^2}}{k^2+i\epsilon} \, .
\ee

\section{Direct Minkowski space calculation}

In general  QFT the Greens functions are not  real functions but complex tempered distributions.
In perturbation theory these are just real poles (together with its Feynman $\ep $ prescription) of the propagators, which when coincide in the loop integrals, produce branch cut starting at the usual production threshold. At one scalar loop level, the two propagators make the selfenergy complex above  the point $p^2=(M_1+M_2)^2$, wherein $M_1,M_2$ are the real masses- in fact the positions of these poles. Considering such one loop correction to the propagator itself then   
depending on the masses of the interacting fields, the real propagator pole persists   when  situated  below the threshold or we get non-zero width and the free particle becomes resonance with finite lifetime. 

 In strong coupling quantum field theory the mechanism of complexification can be very different (however the mixing of both mechanisms is not excluded). 
Here we simply assume that there is no zero at the inverse of propagator for real $p^2$, thus 
$\ep$ factor is not necessary and we integrate over the hyperbolic angles of Minkowski space directly.
For this purpose we have to consider the propagator function as the complex one for all $p^2$.
 A convenient parametrization of the  complex fermion propagator functions $S_s$ and $S_v$  can be written as
\bea
S_s(x)&=&\frac{B(k)}{A^2(k)k^2-B^2(k)}
\nn \\
&=&\frac{R_B\left[(R_A^2-\Gamma_A^2)k^2-R_B^2-\Gamma_B^2\right]+2R_A\Gamma_B\Gamma_A\,k^2}{D}
\nn \\
&+&i\, \frac{\Gamma_B\left[(R_A^2-\Gamma_A^2)k^2+R_B^2+\Gamma_B^2\right]-2R_BR_A\Gamma_A\,k^2}{D} \, ,
\label{ss}
 \\
S_v(k)&=&\frac{A(k)}{A^2(k)k^2-B^2(k)}
\nn \\
&=&\frac{R_A\left[(R_A^2+\Gamma_A^2)k^2-R_B^2+\Gamma_B^2\right]-2R_B\Gamma_A\Gamma_B}{D}
\nn \\
&+&i\, \frac{\Gamma_A\left[-(R_A^2+\Gamma_A^2)k^2
-R_B^2+\Gamma_B^2\right]+2R_A R_B\Gamma_B}{D}\, \, ,
\label{sv}
\eea

where $R_A,R_B$ $(\Gamma_A,\Gamma_B)$ are the real (imaginary) parts of the functions $A,B$
and the denominator $D$ reads
\be
D=([R_A^2-\Gamma_A^2] k^2-[R_B^2-\Gamma_B^2])^2+4(\Gamma_A R_A-\Gamma_B B)^2 \,.
\ee

When solving the Euclidean space SDEs a very common strategy is to use Lorentz scalar arguments of Lorentz invariant functions as a variable. Hence, in order to   compare most easily between  Minkowski and all possibly  considered Euclidean  spaces, we could use the transformation which leaves the Minkowski spacetime interval 
\be
s=t^2-x^2-y^2
\ee
manifestly apparent thorough the calculations (i.e. $s$ could be a variable of the integral SDEs).   

To achieve this we will use 2+1 dimensional pseudospherical (hyperbolic) transformation of 
Cartesian Minkowski coordinates.
The obstacles followed by Minkowski hyperbolic angle integrals when going beyond $A=1$ approximation restrict us to the Landau gauge wherein the $A=1$ is the exact result in TE space. 
In momentum space our convenient choice of the substitution is the following:

\bea \label{myway}
\int d^3k K(k,p)=&&\int_0^{\infty} dr r^2 \int_0^{2\pi} d \theta  \int_0^{\infty} d \alpha
\left\{\sinh\alpha \, {\mbox{\begin{tabular}{|c|}
$k_o=-r\, \cosh\alpha $ \\ $k_x=-r \,\sinh\alpha \, \sin\theta$ \\
$k_y=-r \,\sinh\alpha\, \cos\theta$ \\
\end{tabular}}}
+\sinh\alpha\, {\mbox{\begin{tabular}{|c|} $k_o=r\, \cosh\alpha$\\ $k_x=r \,\sinh\alpha\, \sin\theta$\\ $k_y=r \, \sinh\alpha\, \cos\theta$
\end{tabular}}}\right. 
\nn \\
&&+\left.\cosh\alpha\, {\mbox{\begin{tabular}{|c|} $k_o=-r \sinh \alpha$\\$ k_x=-r \cosh\alpha\, \sin\theta$\\$ k_y=-r \cosh\alpha\, \cos\theta$
\end{tabular}}}+
 \cosh\alpha\, {\mbox{\begin{tabular}{|c|} $k_o=r \sinh \alpha$\\$ k_x=r \cosh\alpha\, \sin\theta$\\ $k_y=r \cosh\alpha\, \cos\theta$
\end{tabular}}}\right\} \, K(k,p) \, .
\eea

Notice, the integral boundaries are universal for  all the subregions of Minkowski space,
  the first line corresponds to the integration over the timelike 2+1momentum where we have 
\be
k^2=k_o^2-k_x^2-k_y^2=r^2>0 \, ,
\ee
where the left term corresponds to the negative energy interval $k_0<-\sqrt{k_x^2+k_y^2}$
and the right term corresponds to the positive  $k_0>+\sqrt{k_x^2+k_y^2}$. The second line
stands
 for the spacelike regime of the integration
\be  
k^2=-r^2<0\, ,
\ee
where the left term corresponds to the minus energy component interval  $k_0= (-\sqrt{k_x^2+k_y^2},0)$, while the right term in the second line stands for positive $k_0= (0,\sqrt{k_x^2+k_y^2})$ subspace of the full 2+1 dimensional Minkowski space. Function $K$ in Rel. (\ref{myway}) represents any kernel in the SDE.

The functions $A,B$  are Lorentz scalars, thus they can depend on $p^2$ only. We freely take the simple choice  of timelike external momenta as $p_{\mu}=(p,0,0)$ which leads to the following $\alpha$ integrals
 for the timelike part of internal momenta:

\be \label{andula}
\int_0^{\infty} d\alpha  \frac{\sinh \alpha}{r^2+p^2+2pr\cosh\alpha}+
\int d\alpha  \frac{\sinh \alpha}{r^2+p^2-2pr\cosh\alpha}
=\frac{1}{pr}\ln\left|\frac{r-p}{r+p}\right| \, , 
\ee
where without any ambiguities $p=\sqrt{p^2},r=\sqrt{r^2}$.

The contribution stemming from spacelike part of loop momenta gives zero because of negative and positive time volume contributions, although each infinite separately, they cancel each other. Note, the $\theta$ integrals contributes by simple  $2\pi$ prefactor.

Integrating over the angles we can see that at the level of our approximation, the SDE separate for spacelike and timelike regime of the threemomenta. For timelike $p$ we get for the function $B$
\be \label{ET1}
B(p^2)=m+i(2+\xi)\frac{e^2}{4\pi^2}\int_0^{\infty} dk
\frac{k}{p}\ln\left|\frac{k+p}{k-p}\right|S_s(k) \, ,
\ee
where $\xi$ is a gauge parameter (and where we trivially changed $r$ of (\ref{andula}) to $k$)
Stressed here, the Eq. (\ref{ET1}) is derived without any requirement of analyticity for the the propagator and the kernel.

For  the external spacelike Lorentz three-vector of momenta the $\alpha$ integration over the spacelike regime gives zero. Taking for instance $p_{\mu}=(0,p,0)$ this can be most easily seen by the inspection of the integrals
\bea
&&\int_0^{\infty} d\alpha  \frac{\cosh \alpha}{-r^2+p^2+2pr\cosh\alpha\sin\theta}+
\int d\alpha  \frac{\cosh \alpha}{-r^2+p^2-2pr\cosh\alpha\sin\theta}
\nn \\
&&=-\frac{a}{b}\int_0^{\infty} d\alpha\left(\frac{1}{a+b\cosh\alpha}-\frac{1}{a-b\cosh\alpha}\right)
=0
\label{iclka}
\eea  
where $a=p^2-r^2\, ,b=2pr\sin\theta$ (and we take $r=\sqrt{r^2},p=\sqrt{-p^2}$) The  integrals in bracket (\ref{iclka}) can be evaluated by using of the following formula:
\bea \label{tabule}
&&\int_0^{\infty} d\alpha\frac{1}{a+b\cosh\alpha}=
\frac{1}{\sqrt{a^2+b^2}}\ln \frac{a+b+\sqrt{a^2-b^2}}{a+b-\sqrt{a^2-b^2}} \,
\nn \\
&&\int_0^{\infty} d\alpha\frac{1}{a+b\sinh\alpha}=
\frac{1}{\sqrt{a^2+b^2}}\ln \frac{a+b+\sqrt{a^2+b^2}}{a+b-\sqrt{a^2+b^2}}
\eea
 
The remaining contributions stem from the combination of the external spacelike  and the internal timelike momenta, putting  now $a=p^2+r^2\, ,b=2pr\sin\theta$ we can get 
\bea \label{beruska}
&&\int_0^{\infty} d\alpha  \frac{\sinh \alpha}{r^2+p^2+2pr\sinh\alpha\sin\theta}+
\int d\alpha  \frac{\sinh \alpha}{r^2+p^2-2pr\sinh\alpha\sin\theta}
\nn \\
&&=-\frac{a}{b}\int_0^{\infty} d\alpha\left(\frac{1}{a+b\sinh\alpha}-\frac{1}{a-b\sinh\alpha}\right)
=\frac{2}{b}\frac{\ln\left(\sqrt{1+(b/a)^2}-b/a\right)}{\sin\theta \,\sqrt{1+(b/a)^2}}
\eea  
where the definite integrals in the second line are given by second line in Eq. (\ref{tabule}) and  again $p,r$ are real and positive variables $p=\sqrt{-p^2},r=\sqrt{r^2}$ .

Adding the all together we get for $p^2<0$ (for $\xi=0$ and $A=1$) the following integral expression for the function $B(p^2)$:
\bea  \label{surprise}
&&B(p^2)=m+i\frac{e^2}{(2\pi)^3}\int_0^{\infty} dr \int_0^{2\pi} d\theta \frac{r}{\sqrt{-p^2}}
\frac{\ln\left(\sqrt{1+z^2}-z\right)}{\sin\theta \,\sqrt{1+z^2}}
\frac{B(r^2)}{r^2-B^2(r^2)} \, ,
\nn \\
&&z=\frac{2\sqrt{-p^2} r \sin\theta}{r^2+p^2}  \, ,
\eea

Let us  stress the main difference when compared to the standard  treatment. Here, this is the timelike part of Minkowski subspace where the results are most naturally obtained.
Quite opposite to the standard approach where Minkowski solution is constructed by continuation of the Euclidean result, here the solution for the spacelike argument is non-trivially made from the timelike solution which must be found as a first. Actually, the solution $B(r^2)$ at the timelike domain $r^2>0$  given by Eq. (\ref{ET1}) is needed in the rhs. of Rel. (\ref{surprise}).

As time is passing we can conclude this Section by the  note that the same results has  been already derived in a bit different manner. The detailed derivation of the Eq. (\ref{ET1}) without the explicit use of the hyperbolic coordinates has been recently presented in \cite{SABA2009}. In the paper \cite{SABA2009} it is also suggested that this is just the solution for timelike subspace  which can be advantageously used to construct the full Minkowski space solution.

\section{Numerical solutions for timelike $p^2$}

Usual approach of obtaining timelike solution is to use the reconstruction theorem \cite{SW1980,GJ1981}, which rests on the primacy of a Euclidean formulation. The solution for $p^2>0$ momenta  is obtained as an analytical continuation of the solution previously obtained in Euclidean space.  
Before describing our result, it is noteworthy to mention such numerical study of complex singularities \cite{MARIS1995} founded when  the momentum was continued into the complex plane. The author of study \cite{MARIS1995} rotated the integration contour and solved  QED2+1 SDEs system for complex variables selfconsistently (more specifically, the  rays $p e^{i\phi}$ with  constant phase $\phi$ was considered for the external and the internal variables). The location of the singularity was obtained by the extrapolation, since  near the vicinity of complex pole the numerical  procedure was unstable. Recall here, the phase $\pi/2$ would transform standard Euclidean SDE into the Temporal space used here. As  will be shown, we have found the equation stable for this phase, even when the pole is expected near the  real timelike axis. For classically massless case, $m=0$ case, the observed location  of singularity is characterized by $\phi\simeq \pi/4$ and $|p|\simeq M(0)$ where $M(0)$ is the Euclidean infrared mass. The author did not continue the solution further towards timelike axis, because of crossing the singularity, however as we opposed above,  doing opposite could be the correct procedure. One should start from the Temporal Euclidean solution, which being identified with the physical timelike Minkowski solution, the continuation  to the spacelike regime could be performed afterwords.

In this paper we present Minkowski solution of Eq. (\ref{ET1}) for the  mass function $M=B$.  We do not perform the  continuation to the spacelike axis, which remains to be done in the future. We will consider the nonzero Lagrangian mass  $m$ and interaction strength characterized by  charge $e$.
QED2+1 is  a superrenormalizable theory and  as we have neglected the photon polarization
it turns to be completely ultraviolet finite and  no renormalization is required at all.
 We assume that the imaginary part of the mass function is dynamically generated thus in order to get the  numerical solution  we split the SDE (\ref{ET}) to the coupled equations for the real and imaginary part of $B$ and solve these  two coupled 
integral equations simultaneously  by the method of  iterations.

To characterize complex mass solutions it convenient to introduce dimensionless parameter
\be
\kappa=\frac{e^2}{m}.
\ee
We set up the scale by taking $m=1$ at any units. Pure dynamical chiral symmetry breaking $(m=0)$ is naturally achieved  at limiting large  $\kappa$.  In small $\kappa$ limit the perturbation theory could be applicable and we expect the mass function receives only tiny imaginary part, however as we can see it is not in full accordance with perturbation theory discussed in the next section.

\begin{figure}
\centerline{\epsfig{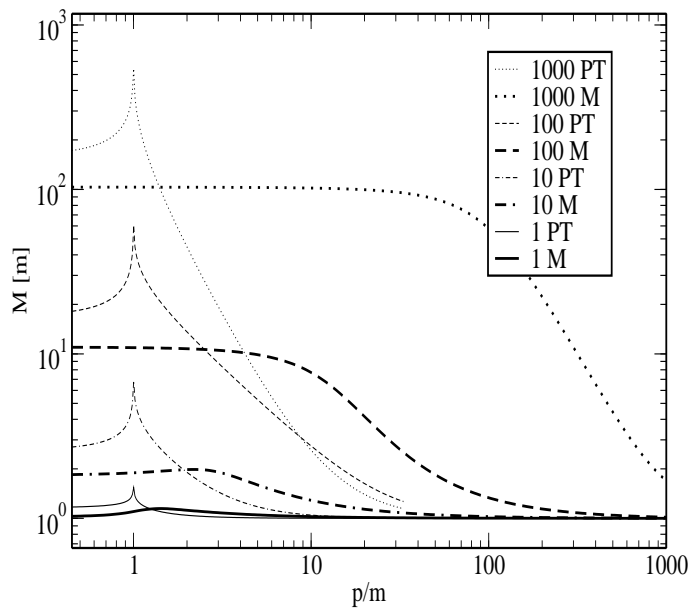}}
\caption[caption]{Magnitude $|M|$ of the running mass $M=|M|e^{i\phi}$ of electron living in 2+1 dimensions for different $\kappa$, scale is $m=1$. 1-loop perturbation theory results on $|M|$ are added for comparison.} \label{velqed}
\end{figure}

The phase $\phi_M$ of the complex mass function defined by $M=|M|e^{i\phi_M}$
is shown in Fig. \ref{uhlyqed} . for a various value of $\kappa$.  For very large $\kappa$ we get the dynamical chiral symmetry breaking in which case the obtained infrared phase  is $\phi_M(p^2=0,\kappa \rightarrow \infty)\rightarrow \pi/2$ while more interestingly it vanishes for very small $\kappa$. There is no imaginary part generation for  fermion selfenergy bellow some "critical" value of $\kappa$ , especially one can observe $\phi_M(p^2=0,\kappa<0.0191\pm 0.0001)=0$ for one flavor ladder QED2+1 (note that the classical mass $m$ represents the full solution with one promile accuracy in the limiting case).    The absolute value of $M$ is displayed in Fig.\ref{velqed} for a various value of the coupling $ \kappa$. As $\kappa$ decreases the expected complex singularities gradually moves from complex plane to the real axis and the function develops  something like very washout threshold enhancement. So for heavy electron we could admit a real pole, albeit confinement.

Confinement of QED2+1 electron  means that we have no free electrons in asymptotic states of any considered process. In quantum theory, physical degrees of freedom are necessarily subject to a probabilistic interpretation implying unitarity and positivity; the physical part of the state space of QCD should be equipped with a positive
(semi-)definite metric. Therefore one way to investigate whether a certain degree of freedom
is confined, is to search for positivity violations in the spectral representation of the
corresponding propagator. 
The standard way in Schwinger-Dyson QCD and QED2+1 studies is to construct  the Schwinger function and check the violation of reflection positivity indirectly from the Euclidean solution \cite{ALKSME2001}. Such  implications of confinement to singularity structure  have been recently studied in the papers \cite{ALDFM2004} in QCD and 
 \cite{BARACORO2008,BRSR2009} in QED2+1. The sufficient condition for the confinement of electrons is the absence of Khallen-Lehmann representation of electron propagator.

Complex mass generation observed assure not only the absence of Khallen-Lehmann representation, but  it has an advantageous technical consequence. The associated absence of the real singularity at $p^2$ axis ensures that, for all $p^2$, the integrand is regular and hence the integral of SDE can be evaluated using straightforward Gaussian quadrature technique, i.e. there are no endpoint or pinch singularities on integration axis,  except  when the coupling constant is very small when compared to the bare mass $m$. In the later case the quantum corrections become negligible and the pole becomes approximately real, however the propagator stays without spectral representation. In all studied cases, it is obvious that the obtained imaginary and real parts of propagators are not related through the Lehmann representation. 
\begin{figure}
\centerline{\epsfig{figure=phase.eps,width=9truecm,height=8truecm,angle=0}}
\caption[caption]{Phase $\phi $ of the dynamical mass function $M=|M|e^{i\phi}$ of electron in QED2+1 for various  $\kappa$, scale is $m=1$.} \label{uhlyqed}
\end{figure}

\section{(In-)Equivalences of  Minkowski, Euclidean and Temporal Euclidean spaces}

The validity of standard Wick rotation is highly speculative topics in the literature. Without large doubts it provides technique which lead to meaningful  solutions of quantum filed theory.  Those results obtained for the Standard Model  are simply testable and actually  
their high accuracy agreement with the experimental observables justify  the assumption made
especially  whenever the perturbation theory framework is applicable.  Assuming the Wick rotation is valid beyond perturbation theory, we could get the Minkowski solution equal to the Euclidean one at spacelike domain of momenta. Here, in our case of 2+1 dimensional confining theory  we know the solution directly in Minkowski space and  the comparison with the Euclidean solution is straightforward and easy task. 

In opposite to  general belief we have  explicitly shown in the previous Section, that the mass function $B$ can be  complex  for all real timelike $p^2$ at large window of the parameters $m$ and $e$. Recall here that Euclidean space Generating functional produces Greens functions which are real by construction. Hence the Minkowski solution $B(0)$ being complex, it is not the one literally known from the Euclidean studies, where $B_E(0)$ is  always purely real number. 
One can speculate wether this feature appears in other strong coupling quantum field theory especially wether this complexification phenomena happen to QCD Greens function at scales $\simeq \Lambda$.  In this case the Minkowski solution cold not be the analytical continuation of usually assumed homomorphic Schwinger function,  but perhaps it is given by some analogue of Eq. (\ref{surprise}) derived here for our simple model. Above speculation and its possible impacts on SDE QCD  phenomenology attracts its own attentiton.

For purpose of completeness and for certain comparison we present and further discuss the various solutions based on the use of different spaces where the quantum field theory can be defined, e.g.
we discuss briefly the perturbation theory result and the solution of SDE obtained in standard
(spacelike) Euclidean space.  Furthermore, in the ladder approximation  we are able to  exhibit the equivalence of Minkowski QED2+1 with QED3 defined in the Temporal Euclidean $(TE)$ space, where the metric simplifies and usually allows to consider more complicated approximations of SDEs. Comparison with perturbation theory and spacelike Euclidean solution will follow afterwords.

\subsection{Proof of equivalence of  ladder QED  formulated in Minkowski and Temporal Euclidean space.}

In the paper \cite{SAZO2008} it was proposed that N-dimensional analog of Wick rotations
performed for space components of Minkowski $N+1$-vector can be partially useful for  the study of a strong coupling quantum field theory.
In even dimensional  3+1QCD it is  the  nonperturbative mechanism  (there is no imaginary unit $i$ presented in the measure)  which is fully responsible for the complex mass generation phenomena. 
In very similar fashion to QED2+1 , it was shown this complexity is responsible   for the absence of real pole type singularities in propagator and hence for confinement. In this Section we will show  that the Temporal Euclidean formulation of our considered SDE exactly agrees with the one derived  in Minkowski space.

In odd dimensional theory, like QED2+1 we study here, the complexification of masses and couplings can be quite naturally expected  because presence of  $i$  in the measures of the integrals defined in TE space. 
To see this explicitly, let us consider the momenta as complex variables and let us  assume that there are no singularities in the second and the fourth quadrants of complex planes of  $k_x,k_y$ in the momentum integrals (see \ref{gap}).  Deforming the contour appropriately then the aforementioned generalized Wick rotation  gives the following prescription for the momentum measure:
\bea
&&k_{x,y} \rightarrow i k_{2,3} \, ,
\nn \\
&&i \int d^3k \rightarrow  -i\int d^3k_{TE} \, ,
\eea
 which, contrary to our standard  $3+1$ space-time, leaves the additional $i$ in front. 

In TE space  the singularity of the free propagator remains,   
for instance the free propagator of scalar particle is
\be
\frac{1}{p^{2}-m^2+\ep} \, ,
\ee 
with a positive square of the  three-momenta 
\be
p^2_{TE}=p_1^2+p_2^2+p_3^2 \, ,
\ee
 thus formulation of the weak coupling (perturbation) theory, albeit possible, would not be more helpful then the standard approach (Wick rotation).

The advantage of the transformation to TE space becomes manifest, since the  fixed square Minkowski momentum $p^2=const$ hyperboloid with infinite surface is transformed into the finite  3dim-sphere in TE space. The Cartesian variables are related to the  spherical coordinates as usually:
\bea
&&k_3=k \cos\theta
\nn \\
&&k_1=k \sin\theta \cos \phi
\nn \\
&&k_2=k \sin\theta \sin \phi \, .
\eea

Making  aforementioned 2d Wick rotation, taking the Dirac trace on $\Sigma$ and integrating over the angles we get for the function $B$
\be \label{ET}
B(p^2)=m+i(2+\xi)\frac{e^2}{4\pi^2}\int_0^{\infty} dk
\frac{k}{p}\ln\left|\frac{k+\sqrt{p^2}}{k-\sqrt{p^2}}\right|S_s(k^2) \, ,
\ee
which is the same equation we derived directly in Minkowski space (\ref{ET1}).

Making the trace  $ Tr \not p\Sigma/(4p^2)$ and integrating over the angle we get for the renormalization wave function in TE space
\bea \label{ET2}
A(p)&=&1+i\frac{e^2}{4\pi^2}\int_0^{\infty} d k
\frac{k^2}{p^2} S_v(k^2) \left[-I+(1-\xi)I\right]
\nn \\
I&=&1+\frac{p^2+k^2}{2\sqrt{p^2 k^2}}\ln\left|\frac{k-\sqrt{p^2}}{k+\sqrt{p^2}}\right|
\eea
with the propagator function $S_v$ defined by (\ref{sv}).
The first term in the bracket $[]$ follows from the metric tensor while the second one proportional to
$(1-\xi)$  stems from the longitudinal part of gauge propagator.
We can see that like in the standard Euclidean  formulation  we get $A=1$ exactly  in quenched rainbow approximation in Landau gauge $\xi=0$.

\subsection{Comparison with the standard treatments}

Phenomena of dynamical chiral symmetry breaking and confinement lie beyond realm of perturbation theory. It is the electron charge $e^2$ which is an intrinsic dimensionful parameter in QED2+1 and thus plays quite similar role as $\Lambda_{QCD}$ in four-dimensional QCD. Contrary to perturbative estimate, the nontrivial mass is generated for any value of $e$ , even when setting  $m=0$ explicitly (recall that $\Sigma\simeq m$ in perturbation theory). Within a certain care (as will be discussed bellow $\Sigma$ is allways too large  for $p\simeq m$), the perturbation theory is applicable  in the limit $e^2/m<<1$, while it is  misleading when $e^2/m>1$. It is not only expected  failure of perturbation theory prediction, but  the arguments and results presented in the previous Sections suggest that  assumed validity of  Wick rotation, which otherwise leads to equivalence of Euclidean and Minkowski space calculations, is not justified in QED2+1. 

We have already shown the equivalence between Temporal Euclidean and Minkowski space calculation which appears to be exact one for the ladder approximation of  electron SDE. Since in wasting majority of existing literature the equivalence of (standard) Euclidean and Minkowski space calculation is assumed, it could be  instructive to show the (in-)equivalence for various fractions of $e^2/m$ explicitly. For this purpose we consider 1-loop perturbation theory of massive QED2+1 wherein 
standard Euclidean-Minkowski continuation is assured by the construction. We also solve the (space) Euclidean SDE where we can compare the infrared mass with our Minkowski results. In order to have perturbation treatment meaningful we always consider mass term in classical QED2+1 Lagrangian. The zeroth order "free" propagator reads 
\be 
\frac{1}{\not p-m}\, ,
\ee
thus the real mass $m$ enters  perturbative calculation at any order of $e$.

As a first we discuss perturbation theory (one can  say perturbation solution of SDE) at order $e^2$. The  on-shell value of momenta $p^2=m^2$, corresponding originally with the single pole of the propagator,  turns to be the inverse logarithmic branch point at the one loop level. The appropriate contribution to the mass function in Landau gauge can be written in the following form of the dispersion relation:
\be \label{disrel}
B(p^2)=m-\int_{0}^{\infty}d \omega\frac{ \rho_B(\omega)}{p^2-\omega+\ep}\, ,
\ee
where the absorptive  part of $B$ reads
\be
\pi \rho_B(\omega)=\frac{e^2 m}{4}\theta(\omega-m^2)\frac{1}{\omega^{\frac{1}{2}}}.
\ee
In words: the mass as a function of momenta is real up to the perturbative threshold where the
imaginary part arises discontinuously (recall, it is in opposite to what happens in QED3+1, where the absorptive part continuously starts from zero).

Evaluating the integral (\ref{disrel}) we get  \cite{NIC1978,DOS2000}:

\be
B(p^2)=m+\frac{e^2 m}{2\pi}\left[\frac{1}{2\sqrt{p^2}} ln\left|\frac{m+\sqrt{p^2}}{m-\sqrt{p^2}}\right|+\frac{i \pi}{2\sqrt{p^2}}\right]
\ee
for timelike $p^2$, while for the spacelike $p^2$ we get
\be
B(p^2)=m+\frac{e^2 m}{2\pi} \frac{1}{\sqrt{-p^2}}arctg\frac{\sqrt{-p^2}}{m^2}\, .
\ee

One loop perturbative results are added in Fig. \ref{velqed}, where the magnitudes are compared  better visualization, while the phases  substantially differ.  Of course, the fact that the phase of mass function decreases   as $\kappa$ decreases was expected, however we have quite trivial results $B\simeq m$ in this case. In the vicinity of the branch point the perturbative results overestimates the SDE solution. The absorptive part starts discontinuously thus the mass function is logarithmically  divergent at the threshold, which is a consequence of planar geometry and makes the  estimate of analytical behaviour (e.g. naive definition of propagator residuum) impossible. In fact,  the function  $B$ is arbitrarily large at vicinity of $m$, $B>>m$ very near of $m$, the perturbative theory fails even for arbitrarily small $\kappa$. Perturbation theory is intangible for a correct determination of propagator singularities, so one should say that the appearance of confinement in QED2+1 has a topological origin rather then being related with the strong coupling.

\begin{figure}
\centerline{\epsfig{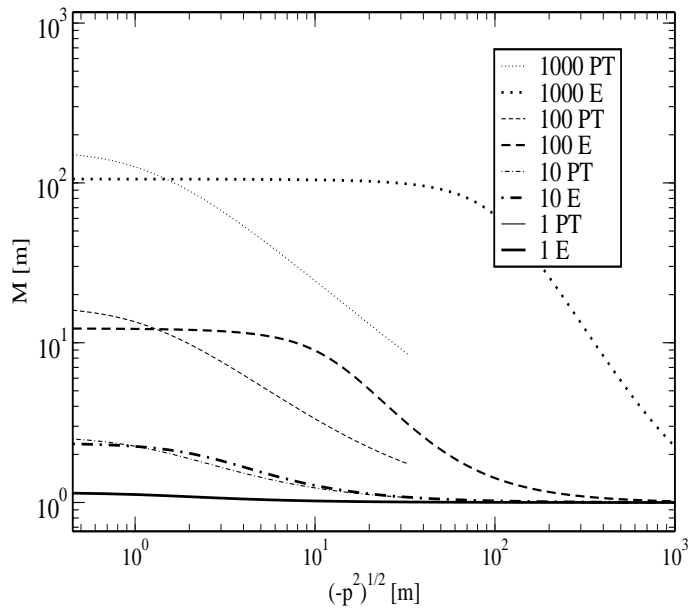}}
\caption[caption]{Comparison of Euclidean solutions and 1-loop perturbation theory for various $e^2/m$.} \label{euclid}
\end{figure}

On the other side, apart of  the threshold region we get reasonable agreement for small $\kappa$ , while increasing $\kappa$ the perturbative obtained mass function always overestimate at infrared while largely underestimate at large $p^2$.   We can observe  the same large $\kappa$ splitting when comparing the perturbative results with the standard Euclidean treatment. The comparison of Euclidean solution and 1-loop perturbative results is shown in Fig. \ref{euclid}. 
 For this purpose we have solved ladder SDE in standard Euclidean space. The equation we have actually solved numerically reads:

\be \label{ES}
B_E(p_E^2)=m+\frac{e^2}{2\pi^2}\int_0^{\infty} dk_E
\frac{k_E}{p_E}\ln\left|\frac{k_E+p_E}{k_E-p_E}\right|S_s(k^2_E) \, ,
\ee
where now $k_E=\sqrt{k^2_E}=\sqrt{-k^2}$, $p_E=\sqrt{p^2_E}=\sqrt{-p^2}$. 

As expected,  there is a slightly better agreement in between the Euclidean SDE solution and  spacelike perturbation theory and for instance for $\kappa=1$ the lines are nearly indistinguishable. However , as in the case of timelike regime we can see the same effect for larger $\kappa$. Perturbation theory mass function is overestimated in the infrared and underestimated for large momenta when compared with the Euclidean SDE solution, of course, now the both solutions are real.

\begin{figure}
\centerline{\epsfig{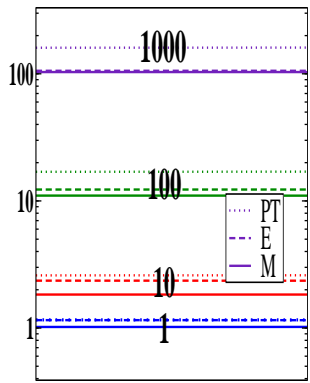}}
\caption[caption]{Magnitudes of dimensionless infrared dynamical mass $|M(0)|/m$ as calculated in various frameworks, solid lines represent Minkowski space ladder SDE solution, dashed lines stand for the  Euclidean space solutions and the dotted lines represent 1-loop perturbation theory with the bare mass $m=1$ in any units. These three lines corresponding to the same coupling are labeled by the appropriate value of dimensionless quantity $e^2/m$.} \label{mzerp}
\end{figure}

More interestingly, keeping the both -the  timelike and Euclidean  (spacelike) solutions we can compare them at the matching point $p^2=0$. This comparison is shown in Fig. \ref{mzerp}. We can see that the agreement (disregarding the phase again) is always better then in the case of comparison with perturbation theory. Furthermore, the solutions start to match again for large $\kappa$, since the negative interference in Minkowski SDE is getting suppressed. For large $\kappa>>1$ we get very easy prescription for the continuation between Minkowski and Euclidean space:

\be
B(p^2)=i B_E(p^2_E\rightarrow p^2)\, \, , \, for \, \, \kappa\rightarrow\infty \,  , 
\ee 
which is true for all $p<e^2$. However , there is no any reason to expect this "symmetry" beyond the ladder  approximation considered here.

\section{Summary and conclusions}

We have presented for  the first analysis of the  electron gap equation  
in  Minkowski and Temporal Euclidean space. The  dynamical generation of  imaginary part of the fermion mass  leads to the absence of Khallen-Lehmann representation, providing thus confining solution for all value of $m$. Apart very small $\kappa$ the real pole in the propagator is absent as well. Similarly to Euclidean QED3 Minkowski QED2+1  exhibits spontaneous chiral symmetry breaking -the mass function  has nontrivial solution in the limit $m=0$, however the mass is complex function.

Furthermore, we compare with QED solved in similar approximation in spacelike Euclidean and Temporal Euclidean space. As a interesting results, although based on the simple
ladder approximation, is the proof of the exact equivalence between  the theories defined in Minkowski 2+1 and 3D Temporal Euclidean space. We expect large quantitative changes when the polarization effect is taken account, especially the existence of critical number of flavors can be different when compared to the known Euclidean space estimates \cite{BARACORO2008,BRSR2009}. 
Opposite to naive belief we showed and explained  in the Section V. that the Wick rotation -the well known calculational trick in quantum theory- provides continuation of Schwinger  function of the Euclidean theory which do not correspond with the Greens function calculated directly in the original Minkowski space. 

We can note our  finding has a little to do with the know   usefulness of various extensions (largely anisotropic and nonrelativistic) of QED3 as an effective theories of the pseudogap insulator/superconducting  phase transition \cite{HERBUT2002,HELE2003} . The finite temperature QED3 Euclidean action seems to be enough for the effective  description of the phenomena. However the inequivalence between Euclidean and spacelike Minkowski subspace should be kept mind  when trying to make some conclusion based on the naive continuation the real  time metric.

More interestingly, one can expect new  outcomes when the similar ideology is applied to the strong coupling even dimensional theories, e.g. to the most successful theory of the strong interaction in the nature: QCD. Note, similar strategy in renormalizable theories
is not so straighforward because of presented infinities. In this case, following the experience from QED2+1 here and proposing an approximations  in Temporal Euclidean space  \cite{SABA2009b} can offer a simple guide which can be realistic enough to provide reasonable Minkowski space SDEs framework for low energy hadronic physics.

\begin{center}{\large \bf  Acknowledgments}\end{center}

We would like to V. Gusynin for useful comments. This work has been financially supported by FCT Portugal, CFTP IST Lisbon.


\end{document}